\documentclass[preprint,nofootinbib,12pt]{revtex4-1}

\usepackage[utf8]{inputenc}

\usepackage{fullpage}

\usepackage{subfigure}

\usepackage{amsmath}
\usepackage{amssymb}
\usepackage{latexsym}

\usepackage{boxedminipage}

\usepackage{ifpdf}

\usepackage{slashed}

\usepackage{tikz}
\usetikzlibrary{positioning}
\usetikzlibrary{decorations.pathmorphing}
\usetikzlibrary{decorations.markings}

\newcommand{\gev}{\, \text{GeV}}

\newcommand{\hc}{{\rm h.c.}}

\newcommand{\beq}{\begin{equation}}
\newcommand{\eeq}{\end{equation}}
\newcommand{\beqa}{\begin{eqnarray}}
\newcommand{\eeqa}{\end{eqnarray}}

\newcommand{\lsim}{\mathrel{\rlap{\lower4pt\hbox{\hskip1pt$\sim$}}
    \raise1pt\hbox{$<$}}}         
\newcommand{\gsim}{\mathrel{\rlap{\lower4pt\hbox{\hskip1pt$\sim$}}
    \raise1pt\hbox{$>$}}}         

\begin{document}

\tikzset{ 
  scalar/.style={dashed},
  scalar-ch/.style={dashed,postaction={decorate},decoration={markings,mark=at
      position .5 with {\arrow{>}}}},
  fermion/.style={postaction={decorate}, decoration={markings,mark=at
      position .5 with {\arrow{>}}}},
  gauge/.style={decorate, decoration={snake,segment length=0.2cm}},
  gauge-na/.style={decorate, decoration={coil,amplitude=4pt, segment
      length=5pt}}
}


\vspace*{.0cm}

\title{A new CP violating observable for the LHC}

\author{Joshua Berger}
\author{Monika Blanke}
\author{Yuval Grossman}

\affiliation{\vspace*{4mm}Institute for High Energy Phenomenology\\
Newman Laboratory of Elementary Particle Physics\\ Cornell University,
Ithaca, NY 14853, USA\vspace*{6mm}}

\begin{abstract}
We study a new type of CP violating observable that arises in three
body decays that are dominated by an intermediate resonance. If
two interfering diagrams exist with different orderings of final
state particles, the required CP-even phase arises due to the different
virtualities of the resonance in each of the two diagrams. This method
can be an important tool for accessing new CP phases at the LHC and
future colliders.
\end{abstract}

\maketitle

\section{Introduction}

The Standard Model (SM) of particle physics contains a single CP
violating phase in the Cabibbo-Kobayashi-Maskawa (CKM) matrix
\cite{Kobayashi:1973fv}. Many different measurements have confirmed
that the SM describes observed CP violation to extremely good accuracy
\cite{Charles:2004jd,Bona:2006sa}. Particularly strong constraints on
physics beyond the SM can be obtained from neutral kaon mixing,
pushing the scale of generic new CP violation operators to at least
$\mathcal{O}(10^5\,\text{TeV})$~\cite{Bona:2007vi}. Strong constraints
have also been obtained from the non-observation of electric dipole
moments (see \cite{Raidal:2008jk} for a review).  While one might be
tempted to conclude that new physics must be CP conserving up to very
high energy scales, this is not necessarily the case.  In fact, it
is sufficient to introduce the new sources of CP violation in such a
way that they are hidden from flavor physics observables. A large new
physics scale is not the only way to achieve this; other options
include, for example, the introduction of flavor symmetries or
decoupling the new sources of CP violation from the flavor sector.
Consequently it is important to search for new physics CP violation
not only indirectly in low energy observables, such as meson decays or
electric dipole moments, but also directly in the production and decay
of new heavy particles at colliders. Direct searches have the
advantage of giving much cleaner access to the new CP violating phases
in question. 

In order to observe CP violation in heavy particle decays, asymmetries
in the decay rates corresponding to CP-conjugate processes can be
measured: 
\begin{equation}
  \label{eq:decay-asymmetry}
  \mathcal{A}_\text{CP} = \frac{\Gamma(M \to f) - \Gamma(\overline{M} \to
    \overline{f})}{\Gamma(M \to f) + \Gamma(\overline{M} \to
    \overline{f})}.
\end{equation}
For this asymmetry to be non-vanishing, the amplitude for the decay
rate must be composed of at least two interfering amplitudes with
different CP-even (``strong'') and CP-odd (``weak'') phases. (If the
momenta, and possibly the 
helicities, of the final state particles can be determined, then it
is possible to avoid the condition of requiring amplitudes with
different strong phases by looking at triple product asymmetries.
See e.\,g.\ Refs. \cite{Valencia:1988it,Korner:1990yx,Kayser:1989vw,Bensalem:2000hq,Datta:2003mj,AguilarSaavedra:2004hu,Bartl:2004jj,Langacker:2007ur,MoortgatPick:2009jy}.) 

In the SM, the weak phase always depends on the CKM phase. More
generally, it is related to complex phases of the Lagrangian parameters
and, therefore, changes sign under CP conjugation.
Strong phases are so-named because they often arise from
strong-interaction rescattering of the final state.  However, several
cases are known where a calculable strong phase arises from the
propagation of intermediate state particles. For instance when the two
amplitudes arise due to mixing of states with the same quantum
numbers, as in $B \to \psi K_S$ for example, the strong phase arises
simply through the time evolution of the intermediate $B^0-\bar B^0$
system. Another source of strong phases is finite width effects that
have been considered in both particle
production~\cite{Pilaftsis:1990mq,Nowakowski:1991si} and
decay~\cite{Eilam:1991yv,Atwood:1994zm}.

The requirement of the existence of a strong phase places a
limitation on our ability to measure CP-violation.  There is no reason
to assume the strong phase is large.  Furthermore, there is often
no way to determine the strong phase for a given process, since it can
involve complicated strongly coupled physics.  It is therefore
important to look for processes where either the strong phase can be
divided out or calculated.  Situations where the strong phase can be
calculated arise most readily in processes involving a propagating
intermediate unstable particle.

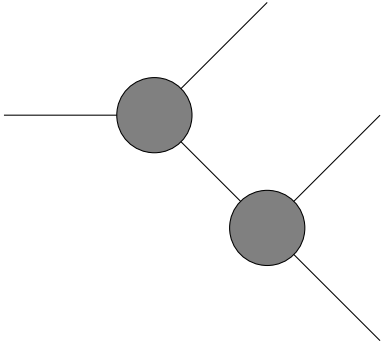
\begin{figure}
  \centering
  \begin{tikzpicture}
    \draw (0,0) -- (2,0) -- (3.5,1.5);
    \draw (2,0) -- (3.5,-1.5);
    \draw (5,0) -- (3.5,-1.5) -- (5,-3);
    \filldraw[draw=black,fill=gray] (2,0) circle (0.5cm);
    \filldraw[draw=black,fill=gray] (3.5,-1.5) circle (0.5cm);
    \end{tikzpicture}
  \caption{Diagram demonstrating the presence of a strong phase in the propagator of an intermediate state.}
  \label{fig:strong-phase-prop}
\end{figure}
In order to see this, consider a diagram of the form shown in
Fig.~\ref{fig:strong-phase-prop}.  The corresponding 
amplitude can generally be written in the form
\begin{equation}
  \label{eq:amp-form}
  \mathcal{M} = \mathcal{M}_1 \frac{1}{q^2 - m^2 + i \Gamma m} \mathcal{M}_2,
\end{equation}
where $\mathcal{M}_{1,2}$ are, roughly, amplitudes for the upper and
lower parts of the diagram and carry the weak phase, $q$ is the
off-shell momentum of the propagating particle, $m$ is its mass, and
$\Gamma$ is its width.  The Breit-Wigner denominator in this amplitude
is CP-even; that is, in the CP-conjugate amplitude, the $i$ in the
denominator appears with the same sign.  Thus, the propagating
particle leads to a strong phase
\begin{equation}
  \label{eq:bw-strong-phase}
  {\rm arg}\left(\frac{1}{q^2 - m^2 + i \Gamma m}\right).
\end{equation}

Recall that in order to have observable CP-violation in a decay process,
it is necessary to have at least two amplitudes with different strong phases.
If both amplitudes have a propagating particle, then there are two
ways in which their strong phases can differ:
\begin{enumerate}
\item The propagating particles could be different, so that they have
  different mass and/or width;
\item The propagating particles could be the same, but off-shell by
  different amounts.
\end{enumerate}

When studying SM physics, the first situation was considered
\cite{Eilam:1991yv,Atwood:1994zm}. In this paper, we study the second
case, that is, strong phases that arise from different
virtualities. Unlike the SM, this effect is a common feature of heavy
particle decays leading to CP violating asymmetries in new physics
models.  In order to obtain a non-vanishing CP even phase the decay
must proceed via two interfering diagrams with the same intermediate
unstable particle but with different orderings of the final
states. The examples studied in this paper deal with neutral
Majorana-like particle decays, i.\,e.\ particles that transform under
real representations of all symmetry groups. While the appearance of a
CP even phase is very natural in such a situation,  the mechanism in
question is also present in charged particle decays.\footnote{We would
  like to thank Alejandro Szynkman for useful discussion that led us
  to this observation.}  A concrete example for the latter within the
Littlest Higgs model with T-parity will be the subject of a
forthcoming publication \cite{BBG2}.

The remainder of this paper is structured as follows.  In Section
\ref{sec:cp-even-phases}, 
we discuss general considerations for having a non-vanishing strong
phase difference as described above, using a toy model for
concreteness.  In Section~\ref{sec:cp-violation-charged}, we present
results of a study of CP violation via that mechanism within a model
of new physics, the Minimally Supersymmetric Standard Model (MSSM).
We discuss these results and conclude in Section
\ref{sec:conclusions}.

\section{CP-even Phases in the Propagator}
\label{sec:cp-even-phases}

As discussed in the introduction, three conditions must be satisfied
in order for a CP-violating asymmetry to be observable in a given process:
\begin{enumerate}
\item the amplitude must be composed of at least two terms $a_1$ and
  $a_2$;

\item the two terms must have different CP-even (``strong'') phases $\delta_1 \neq
  \delta_2$;

\item the two terms must have different CP-odd (``weak'') phases
  $\phi_1 \neq \phi_2$. 
\end{enumerate}
In other words the amplitude must have the structure
\begin{equation}
\mathcal{M} =|a_1|e^{i(\delta_1+\phi_1)}+|a_2| e^{i(\delta_2+\phi_2)}\,.
\end{equation}
The asymmetry $\mathcal{A}_\text{CP}$ defined in (\ref{eq:decay-asymmetry}) is then given by
\begin{equation}
  \label{eq:asym-cond-sat}
  \mathcal{A}_\text{\rm CP} \propto |a_1| |a_2| \sin(\delta_1 - \delta_2)
  \sin(\phi_1 - \phi_2),
\end{equation}
where we see explicitly that the three conditions must be satisfied.  

As discussed in the introduction, two decay amplitudes can have different
CP-even phases if the intermediate propagating particles are off-shell
by different amounts. To make this statement more concrete, consider a
three body decay $X^0_0 \to X_1^+ X_2^- X_3^0$.  Suppose further that
this decay can proceed in two ways
\beq
X_0^0 \to X_1^+ Y^{-*} \to X_1^+
X_2^- X_3^0, \qquad
X_0^0 \to X_2^- Y^{+*} \to X_1^+ X_2^- X_3^0.
\eeq
In both cases, the off-shell particle is $Y$ and clearly has the same
mass and width.  However, its four-momentum in each case is different
for a given point in the available phase space of the decay.  The two
decay modes contribute two different terms to the amplitude which have
different strong phases at this point in phase space.

\begin{figure}[!t]
  \centering
  \begin{tikzpicture}
    \draw[scalar] (0,0) node [left] {$X_0^0$} -- (2,0);
    \draw[scalar-ch] (3.414,1.414) node [right] {$X_i^-$} -- (2,0);
    \draw[scalar-ch] (2,0) -- (3.414,-1.414) node [right] {$Y^+$};
    \node[anchor=west] at (3.9,0) {$= -i a e^{i\varphi_a}$};

    \draw[scalar] (7,0) node [left] {$X_3^0$} -- (9,0);
    \draw[scalar-ch] (10.414,1.414) node [right] {$X_i^-$} -- (9,0);
    \draw[scalar-ch] (9,0) -- (10.414,-1.414) node [right] {$Y^+$};
    \node[anchor=west] at (10.9,0) {$= -i b e^{i\varphi_b}$};
    \end{tikzpicture} \caption{Feynman rules for the toy model.}
    \label{fig:feyn-toy-mod}
\end{figure}
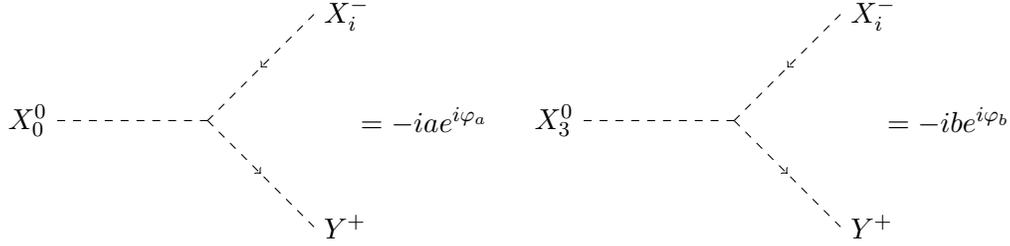

\begin{figure}[!t]
  \centering
  \begin{tikzpicture}
    \draw[scalar] (0,0) node [left] {$X_0^0$} -- (2,0);
    \draw[scalar-ch] (4.828,0) node [right] {$X_2^\mp$} --
    (3.414,-1.414);
    \draw[scalar-ch] (3.414,-1.414) -- node [above right] {$Y^\mp$}
    (2,0);
    \draw[scalar-ch] (2,0) --(3.414,1.414) node [right] {$X_1^\pm$};
    \draw[scalar] (3.414,-1.414) -- (4.828,-2.828) node [right]
    {$X_3^0$};
     \draw[scalar] (7,0) node [left] {$X_0^0$} -- (9,0);
     \draw[scalar-ch]  (10.414,1.414) node [right] {$X_2^\mp$} --
     (9,0);
     \draw[scalar-ch] (9,0) -- node [above right] {$Y^\pm$}
     (10.414,-1.414);
     \draw[scalar-ch] (10.414,-1.414) -- (11.828,0) node [right]
     {$X_1^\pm$}; 
    \draw[scalar] (10.414,-1.414) -- (11.828,-2.828) node [right]
    {$X_3^0$};   
  \end{tikzpicture}
  \caption{Diagrams for the decay $X_0^0 \to X_1^\pm X_2^\mp X_3^0$.}
  \label{fig:diag-toy-mod}
\end{figure}
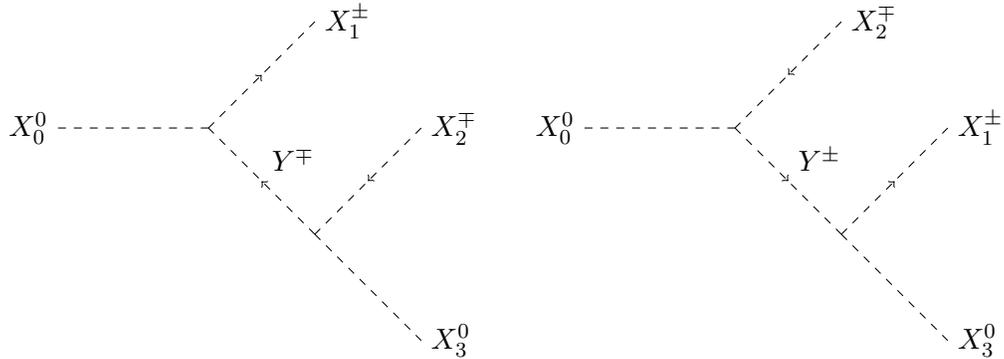

To demonstrate the new CP-even phase, we consider a simple toy model.
We assume that all the particles involved are scalars and consider
only cubic couplings.  We further assume a universality of couplings:
the $X_0^0$ and $X_3^0$ each couple to the charged particles with the
same couplings. While this simplifying assumption is not a  necessary condition, it is crucial that all four couplings of $X_{1,2}^\pm$ to $X_0 Y^\mp$ and $X_3 Y^\mp$ be non-vanishing so that two interfering diagrams with different final state orderings are present. The Feynman rules we consider are presented in
Fig.~\ref{fig:feyn-toy-mod}. This toy model has only one physical CP-odd phase,
\beq
\varphi = \varphi_b-\varphi_a.
\eeq
The diagrams we consider are presented in Fig.~\ref{fig:diag-toy-mod}.
The differential decay width can be obtained from the Feynman rules in
Fig.~\ref{fig:feyn-toy-mod} {and reads
\begin{eqnarray}
\frac{d\Gamma}{dq_{13}^2dq_{23}^2}&=& 
\frac{a^2 b^2}{32 (2\pi)^3m_0^3}\times
\frac{1}{(\Delta\hat q_{13}^2)^2+\hat\Gamma_Y^2}\times
\frac{1}{(\Delta\hat q_{23}^2)^2+\hat\Gamma_Y^2}\times 
\nonumber\\&& 
\Big[\left((\Delta\hat q_{13}^2)^2+(\Delta\hat q_{23}^2)^2+2\hat\Gamma_Y^2\right) 
\nonumber\\ &&\quad
+2\cos(2\varphi) \left[\Delta\hat q_{13}^2 \Delta\hat q_{23}^2+\hat \Gamma_Y^2\right]
+2\sin(2\varphi) \hat\Gamma_Y(\Delta\hat q_{13}^2-\Delta\hat q_{23}^2) \Big]
\end{eqnarray}
where $m_0$ is the mass of $X_0$, $\Gamma$
($\overline{\Gamma}$) is the rate for $X_0^0 \to X_1^+ X_2^- X_3^0$
($X_0^0 \to X_1^- X_2^+ X_3^0$) and
\beq
q_{ij}^2 = (p_i + p_j)^2, \qquad
\hat q_{ij}^2={q_{ij}^2 \over m_Y^2}, \qquad
\hat \Gamma_Y = {\Gamma_Y \over m_Y},
\eeq
where $m_Y$ and $\Gamma_Y$ are the mass and width respectively of
$Y^\pm$. It is convenient to parametrize the differential decay width
using
\beq
\Delta \hat q_{ij}^2 = \hat q_{ij}^2-1\,, 
\eeq
since we
will see that the asymmetry will be largest near the point $q_{13}^2 =
q_{23}^2 = m_Y^2$ in phase space.}

The first asymmetry that we calculate is the differential rate
asymmetry before integrating over phase space:
\begin{equation}
  \label{eq:diff-cp-asym-toy-mod}
  \mathcal{A}_{\rm CP}^{\rm diff} = \frac{d\Gamma/dq_{13}^2 dq_{23}^2 -
    d\overline{\Gamma}/dq_{13}^2 dq_{23}^2}{d\Gamma/dq_{13}^2
    dq_{23}^2 + d\overline{\Gamma}/dq_{13}^2 dq_{23}^2},
\end{equation}
It is given by
\begin{equation}
  \label{eq:diff-cp-asym-toy-mod-res} 
\mathcal{A}_{\rm CP}^{\rm diff} =
  \frac{2 \sin(2\varphi) (\Delta \hat q_{13}^2 - \Delta \hat q_{23}^2)
  \hat\Gamma_Y}{2 [1+\cos(2\varphi)] \hat\Gamma_Y^2 + |\Delta \hat q_{13}^2\,
  e^{i\varphi} + \Delta \hat q_{23}^2 \,e^{-i\varphi}|^2}.
\end{equation}
Note that this asymmetry is proportional to the sine of the weak phase
as desired.  Furthermore, it is proportional to $\hat\Gamma_Y (\Delta
\hat q_{13}^2 - \Delta \hat q_{23}^2)$.  When either $\Gamma_Y = 0$ or $q_{13}^2
= q_{23}^2$ the asymmetry vanishes. This factor in the numerator is
proportional to the CP-even phase difference of the two diagrams.  We
thus demonstrate the occurrence of a CP-even phase due to the virtual
$Y^\pm$ being off-shell by different amounts in the two diagrams.

The denominator of the asymmetry (\ref{eq:diff-cp-asym-toy-mod-res})
is minimized when $\Delta \hat q_{13}^2=\Delta \hat q_{23}^2=0$. That
is, when the $Y^\pm$ is on-shell in both diagrams.  The numerator,
however, also vanishes at that point. We thus expect that the points
in phase space where the size of the asymmetry is maximized are near
the point $q_{13}^2 = q_{23}^2 = m_Y^2$, along the line $\Delta\hat
q_{13}^2 +
\Delta\hat q_{23}^2 = 0$ in order to be as far from the situation where
$\Delta\hat q_{13}^2 =
\Delta\hat q_{23}^2$ as possible.  In this simple model, we can determine
the points of maximum asymmetry analytically and obtain a simple
result: the size of the asymmetry is maximized when
\begin{equation}
  \label{eq:max-asym-toy-mod}
  \Delta \hat q_{13}^2 = \pm \hat \Gamma_Y \cot(\varphi), \qquad 
  \Delta \hat q_{23}^2 = \mp \hat \Gamma_Y \cot(\varphi),
\end{equation}
matching our expectation.  This result is modified in more complex
situations.  In particular, when the two interfering diagrams differ
in size, the maximum asymmetry is pushed closer to one of the
resonances.

Perhaps more telling than the asymmetry itself is the significance
of a CP violating signal.  The significance of a Dalitz plot asymmetry in a specific bin is given by  \cite{Bediaga:2009tr}
\begin{equation}
  \label{eq:significance-dalitz}
  \sigma_\text{CP} = \frac{N(i) - \overline{N}(i)}{\sqrt{N(i) + \overline{N}(i)}}.
\end{equation}
This quantity depends on the number of $X_0$ produced, $N$, so that it
cannot be determined without providing further specifications.  The
relative significance of the bins, however, is of interest as it
determines which bins are most important for confirming the existence
of an asymmetry.  These bins are not necessarily the ones with maximum
asymmetry as the differential rate is enhanced near the resonances.
There is a tension between the asymmetry which is largest away from
the line $q_{13}^2 = q_{23}^2$ and the differential rate which is
largest there.  In the specific case we are considering, this
significance can be written as
\begin{equation}
  \label{eq:significance-toy}
  \frac{d\sigma_\text{CP}}{\sqrt{dq_{13}^2 q_{23}^2}} =
  \sqrt{\frac{N}{\Gamma_{X_0}}}  \frac{d\Gamma/dq_{13}^2 dq_{23}^2 - 
    d\overline{\Gamma}/dq_{13}^2 dq_{23}^2}{\sqrt{d\Gamma/dq_{13}^2
    dq_{23}^2 + d\overline{\Gamma}/dq_{13}^2 dq_{23}^2}},
\end{equation}
where $\Gamma_{X_0}$ is the total width of the $X_0^0$.  

\begin{figure}[!h]
  \centering
\mbox{\subfigure{\includegraphics[width=.45\textwidth]{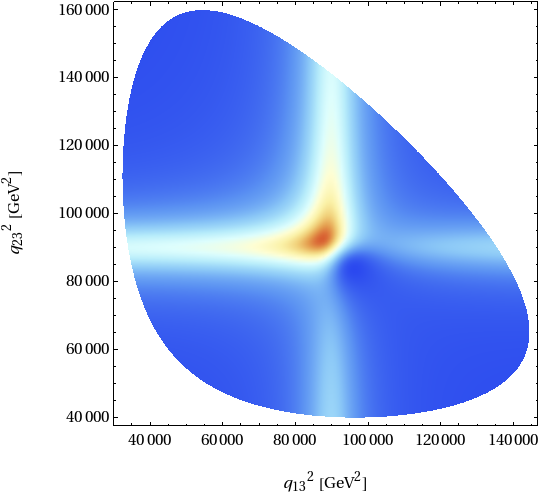}}\qquad
  \subfigure{\includegraphics[width=.45\textwidth]{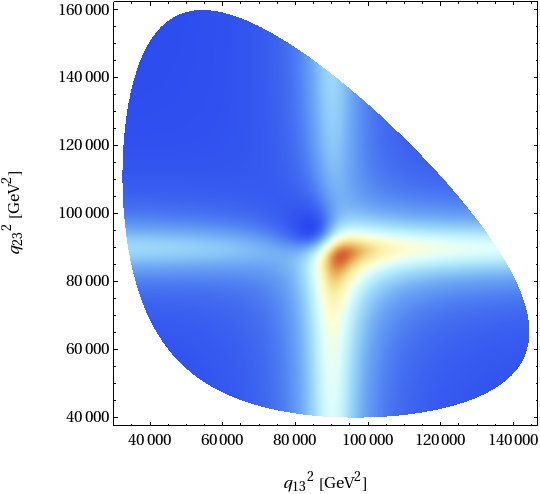} }}

\mbox{\subfigure{\includegraphics[width=.45\textwidth]{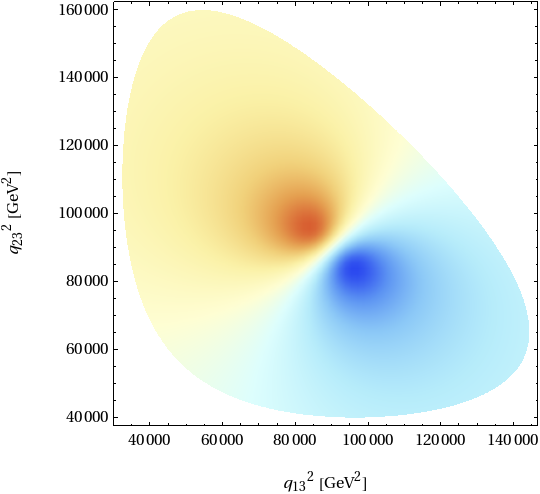}}\qquad
  \subfigure{\includegraphics[width=.45\textwidth]{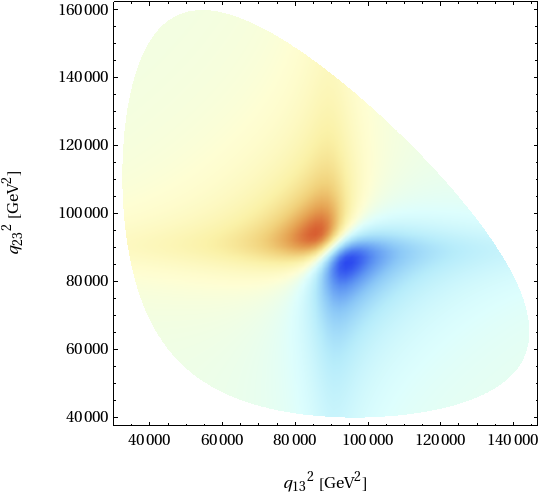}}}
\caption{Dalitz plots for (a) the differential rate of $X_0 \to
X_1^+ X_2^- X_3^0$, (b) the differential rate of the CP conjugate
decay $X_0 \to X_1^- X_2^+ X_3^0$, (c) the asymmetry
$\mathcal{A}_\text{CP}$, and (d) the significance
$\frac{d\sigma_\text{CP}}{\sqrt{dq_{13}^2 dq_{23}^2}}
\sqrt{\frac{\Gamma_{X_0}}{N}}$.}  \label{fig:dalitz-plots-toy}
\end{figure}

The Dalitz plots of the differential rate for $X_0^0$ decay, the rate for the CP conjugate decay, the rate
asymmetry,  and the
significance are given in Fig.~\ref{fig:dalitz-plots-toy}.
To produce that plot we use the following parameters
\beqa
&&m_0= 500\gev ,\quad
m_1=100\gev,\quad
m_2= 120\gev,\quad
m_3= 80\gev,\nonumber \\
&&a=20\gev,\quad
b=30\gev,\quad
\varphi={\pi \over 4}, \quad
m_Y=300\gev ,\quad
\hat\Gamma_Y= 7\%.
\eeqa
All of the features discussed above are observed in these plots.  The
differential rate of $X_0^0$ decay is largest along two resonances at
$q_{13}^2 = m_Y^2$ and $q_{23}^2 = m_Y^2$ with strong interference
where the two resonances overlap.  Interestingly the interference is
constructive above the line $q_{13}^2 = q_{23}^2$ while destructive
below this line.  This feature is reversed for the differential rate
of the CP conjugate decay: now the interference is constructive below
$q_{13}^2 = q_{23}^2$ and destructive above that line, thus exhibiting
a clear sign of CP violation.  This is made even more explicit in
Fig.~\ref{fig:dalitz-plots-toy} (c) showing the differential CP
asymmetry.  The maximum asymmetry is seen to be along the line
$q_{13}^2 + q_{23}^2 = 2 m_Y^2$, but away from the point $q_{13}^2 =
q_{23}^2 = m_Y^2$.  The maximum significance is located closer to the
resonances along the lines $q_{13}^2 = m_Y^2$ and $q_{23}^2 = m_Y^2$.

\begin{figure}[!ht]
  \centering
\mbox{\subfigure{\includegraphics[width=.4\textwidth]{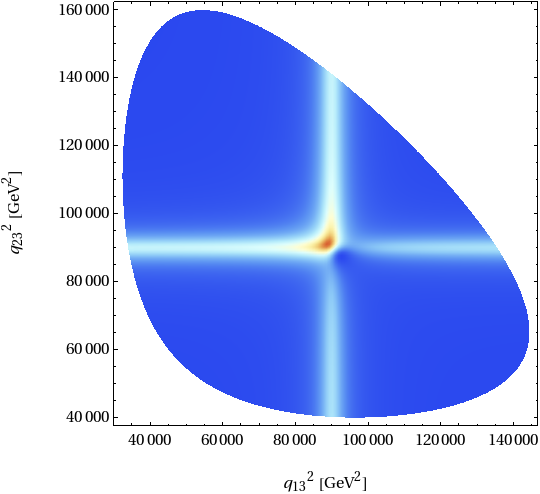}}\hspace{1cm}
  \subfigure{\includegraphics[width=.4\textwidth]{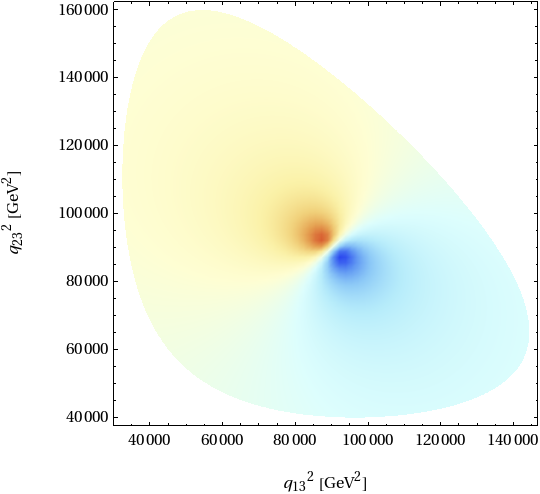} }}

\mbox{\subfigure{\includegraphics[width=.4\textwidth]{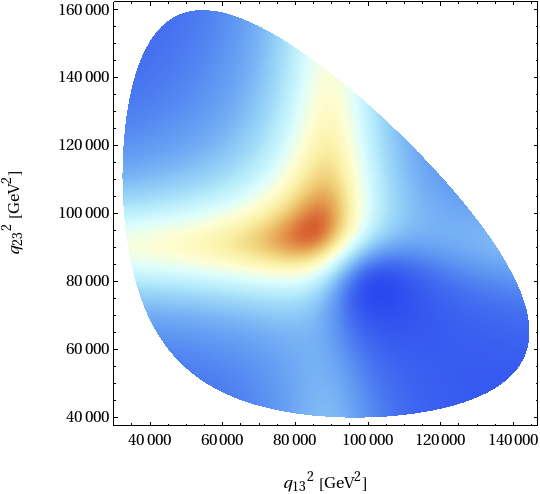}}\hspace{1cm}
  \subfigure{\includegraphics[width=.4\textwidth]{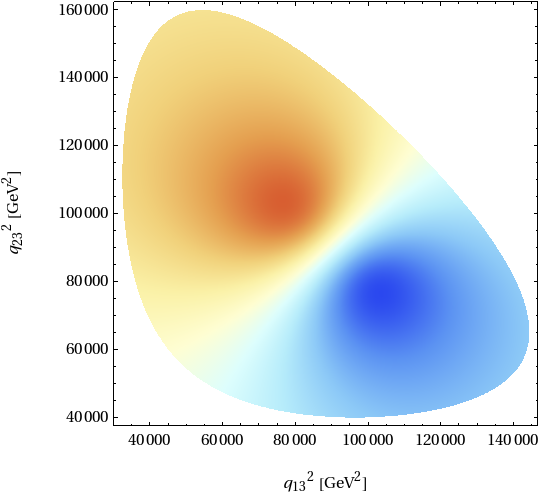} }}

\mbox{\subfigure{\includegraphics[width=.4\textwidth]{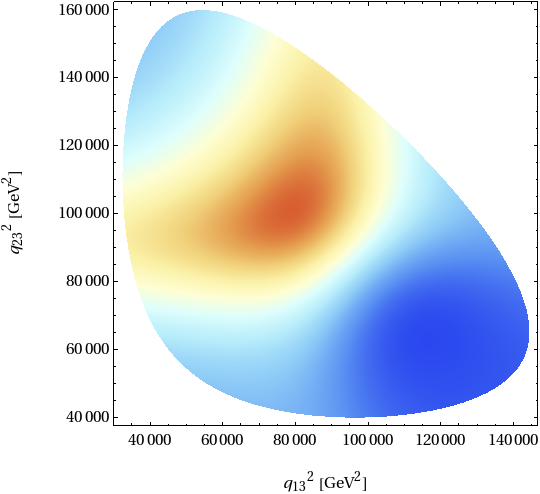}}\hspace{1cm}
 \subfigure{\includegraphics[width=.4\textwidth]{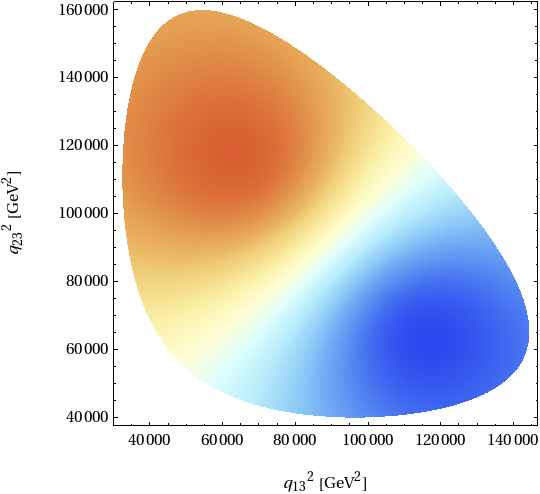} }}

  \caption{Differential decay rate and CP asymmetry for $X_0 \to X_1^\pm X_2^\mp X_3^0$ for $\hat\Gamma_Y= 3\%$, 15\% and 30\% (from top to bottom).}
  \label{fig:Gamma-dependence-toy}
\end{figure}

It is instructive to consider how the obtained results change with the
width of the intermediate state particle $Y^\pm$. The resonances
visible in the Dalitz plot of the differential decay rate should get
broader with increasing $\Gamma_Y$. Consequently also the differential
CP asymmetry is expected to grow and spread further in phase
space. These features are clearly visible from
Fig.~\ref{fig:Gamma-dependence-toy} where we show the differential
decay rate and CP asymmetry for $X_0 \to X_1^\pm X_2^\mp X_3^0$ for
$\hat\Gamma_Y= 3\%$, $15\%$ and $30\%$, in addition to the
corresponding plots for $\hat\Gamma_Y= 7\%$ shown in
Fig.~\ref{fig:dalitz-plots-toy}. We therefore expect the effect in
question to be particularly pronounced in models which predict
strongly coupled resonances.

We now turn to the discussion of integrated asymmetry variables. In
general, it is expected that it will be easier to measure the
integrated asymmetry. We first discuss the total integrated asymmetry
defined as 
\beq\label{eq:intcp}
\mathcal{A}_\text{CP}^{\rm int}
=\frac{1}{\Gamma+\bar\Gamma}\int dq_{13}^2 dq_{23}^2 \left(\frac{d\Gamma}{dq_{13}^2 dq_{23}^2} -\frac{d\bar\Gamma}{dq_{13}^2 dq_{23}^2} \right) \,.
\eeq
Note that the integrated rate asymmetry vanishes in the limit where
the particles $X_1$ and $X_2$ are degenerate.  The asymmetry
(\ref{eq:diff-cp-asym-toy-mod-res}) is anti-symmetric under $q_{13}^2
\leftrightarrow q_{23}^2$, so if phase space is symmetric under such a
transformation, the integrated rate asymmetry vanishes.  In the case
of degenerate charged daughters, phase space has such a symmetry. 
It is therefore beneficial to weigh the asymmetry above and below
$q_{13}^2 = q_{23}^2$ with a relative minus sign and define
\beq\label{eq:weigh}
\mathcal{A}_\text{CP}^{\rm wgt}
=\frac{1}{\Gamma+\bar\Gamma}\int dq_{13}^2 dq_{23}^2 \, \operatorname{sgn}(q_{23}^2-q_{13}^2) \left(\frac{d\Gamma}{dq_{13}^2 dq_{23}^2} -\frac{d\bar\Gamma}{dq_{13}^2 dq_{23}^2} \right) \,.
\eeq
Whether or not this asymmetry can be measured is an experimental
issue.

Even in this simplified model, we are unable to perform a full phase
space integration, due to the rather complex nature of three body
phase space.  We can, however, integrate over a box around the point
$q_{13}^2 = q_{23}^2 = m_Y^2$.  The largest contributions to the rate
asymmetry will come from within such a box.  If the box is a square,
then the region of phase space integration is again symmetric and the
CP asymmetry vanishes.  We could simply use an asymmetric phase space
region, but we gain more sensitivity by taking advantage of the
sign-weighted asymmetry defined in eq.\ \eqref{eq:weigh}.  Integrating
over a square box in phase space with width 
$2 w m_Y^2$
centered at $q_{13}^2 = q_{23}^2 = m_Y^2$ and assuming
that $\hat\Gamma_Y \ll w$, the resulting integrated asymmetry
is given by
\begin{equation}
  \label{eq:int-asym-toy-mod}
  \mathcal{A}_{\rm CP}^{\rm wgt} \approx 
x \log x \sin(2\varphi), \qquad
x=\frac{2\Gamma}{w m} 
\end{equation}
From the Dalitz plot, we conclude that most of the asymmetry effect is
located within such a box.  The full asymmetry will then be of the
same order of magnitude, with $\hat\Gamma_Y \ll w \lesssim 1/4$
required by kinematics.  The asymmetry is then proportional to the
ratio of the width to some combination of mass scales, with a
logarithmic enhancement.  The asymmetry is larger for larger width.

In this section, we have worked with the simplest model that exhibits
CP-violation where the difference in strong phase between the two
diagrams for the process is due to the difference in virtuality of
the off-shell particles.  The model could be complicated by higher
spin particles or by other diagrams.  Independent of these
complications, we can say a few things about the asymmetries.  All of
the asymmetries will of course be proportional to the sine of the weak
phase difference between the diagrams.  The differential rate
asymmetry due to the effects described here will always vanish along
the line $q_{13}^2 = q_{23}^2$.  The integrated rate asymmetry will
always vanish if the phase space is symmetric about $q_{13}^2 =
q_{23}^2$.  By doing a weighted integration over phase space, we can
avoid this last constraint and enhance the asymmetry in cases where
the two charged particles in the final state are nearly degenerate.

Another possible complication that could arise occurs in the large
width limit.  We have worked in the Breit-Wigner approximation, which
will be valid in the new physics scenario we consider below.  If the
intermediate resonance is broad, the Breit-Wigner approximation breaks
down.  This does not alter the qualitative fact that the resonance
virtuality leads to a strong phase. We stress that this generic
feature of unstable modes in any theory is the crucial one for our
purposes.

\section{CP Violation in the Charged Higgs Channel in the MSSM}
\label{sec:cp-violation-charged}

We now turn to study how this new source of CP violation could be
relevant to the MSSM.  The electroweak sector of the MSSM is described
in Appendix~\ref{sec:electr-sect-mssm}. That model is a good starting
point, since in the limit we are considering it contains only one
CP-violating phase, ${\rm Im}(\mu^* b M_2^*)$, defined in
\eqref{eq:cp_vio_invariant}. Any
CP violating observable must involve a process that includes mixing
between the Higgs and the electroweak sectors.  It turns out that the
process
\beq \label{eq:decay}
\chi^0_4 \to \chi^\pm_i \chi^\mp_j \chi^0_1, \qquad i \ne j,
\eeq 
is very instructive for studying the impact of the strong phases
of interest.  This process necessarily involves mixing between the
Higgs and electroweak sectors.  Note that we must be in the limit
where the heaviest neutralino is sufficiently heavy that the decay
(\ref{eq:decay}) is kinematically allowed.  This only occurs when the
$\chi^0_4$ is mostly Bino like and the Bino soft mass $M_1$ is large,
that is
\begin{equation}
  \label{eq:scales-susy}
  m_{\chi^0_4} \sim M_1 \gg m_{\chi^0_i}, m_{\chi^\pm_j} \sim
  \sqrt{|\mu M_2|} > m_Z,
\end{equation}
for $i = 1,2,3$ and $j = 1,2$.  In order for the decay to be
kinematically allowed, the hierarchy must be at least 
\begin{equation}
  \label{eq:kinematic-constraint}
  M_1 \gtrsim 3 \sqrt{|\mu M_2|}.
\end{equation}

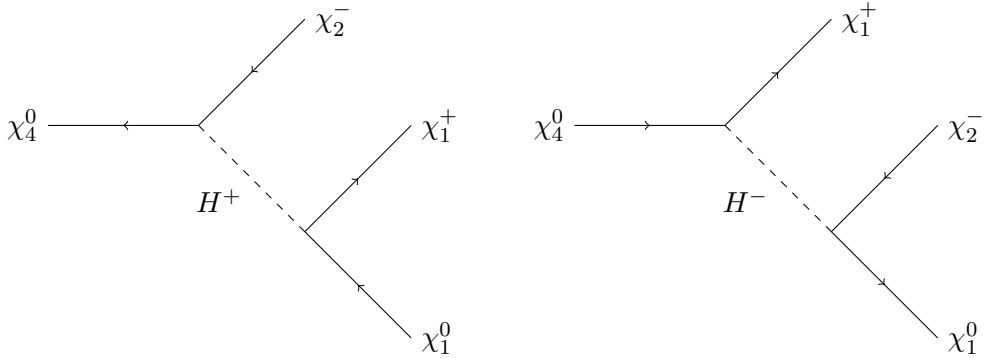
\begin{figure}[!t]
  \centering
  \begin{tikzpicture}
    \draw[fermion] (3.414,1.414) node [right] {$\chi^-_{2}$}-- (2,0);
    \draw[fermion] (2,0) -- (0,0) node [left] {$\chi^0_4$};
    \draw[scalar] (2,0) -- node [below left] {$H^+$} (3.414,-1.414);
    \draw[fermion] (4.828,-2.828) node [right] {$\chi^0_{1}$} -- (3.414,-1.414);
    \draw[fermion] (3.414,-1.414) -- (4.828,0) node [right] {$\chi^+_{1}$};

    \draw[fermion] (7,0) node [left] {$\chi^0_4$} -- (9,0);
    \draw[fermion] (9,0) -- (10.414,1.414) node [right] {$\chi^+_{1}$};
    \draw[scalar] (9,0) -- node [below left] {$H^-$} (10.414,-1.414);
    \draw[fermion] (11.828,0) node [right] {$\chi^-_{2}$} -- (10.414,-1.414);
    \draw[fermion] (10.414,-1.414) -- (11.828,-2.828) node [right] {$\chi^0_{1}$};
  \end{tikzpicture}
  \caption{Decays of a neutralino through a charged Higgs.  In this
    case, there are only two diagrams for the decay.}
  \label{fig:decay-charged-H}
\end{figure}

There are several diagrams for the process $\chi^0_4 \to \chi^\pm_i
\chi^\mp_j \chi^0_1$, but we would like to focus on the diagram
mediated by the charged Higgs as illustrated in
Fig.\ \ref{fig:decay-charged-H}.  We further assume that the charged
Higgs can decay on-shell in both cases, so that $m_{H^\pm} \gtrsim 2
\sqrt{|\mu M_2|}$.  In principle, there are also diagrams 
mediated by the neutral Higgses, the $W$, and the $Z$.  Diagrams with
intermediate $W$ and $Z$ can be neglected since the $W$ and $Z$
are too light to decay on-shell and thus the amplitudes are suppressed
compared to the nearly on-shell amplitudes for the Higgses.  The
lighter neutral Higgs and CP-odd Higgs will also generally be too
light to decay on-shell, but the heavy neutral Higgs will generally
have a mass $m_{H^0} \sim m_{H^\pm}$.  We will, however, neglect all
but the diagrams mediated by the charged Higgs for simplicity.  If
other diagrams were included, then the more familiar type of strong
phase would contribute in the interference between these diagrams and
the charged Higgs mediated ones.

Before performing some analytic and numerical calculations, we would
like to get an idea of how large the CP asymmetry can be in this
case.  To perform this estimate, we take into account the three
sources of suppression that the numerator has relative to the
denominator: the weak phase, the strong phase, and, in the integrated
case, the required phase space asymmetry.  The CP odd effect is
proportional to
\eqref{eq:cp_vio_invariant}, that is to $|\mu b M_2|$. The relevant dimensionless quantity is
normalized to the mass of the decaying particle, that is to some powers
of $M_1$. Taking $b \sim M_1^2$, which is equivalent to taking
$\sin\beta \sim 1$, we conclude that this gives a suppression of
\begin{equation}
  \label{eq:supp-weak-phase} \mathcal{A}_{\rm CP}^{\rm diff} \propto \frac{|\mu
  M_2|}{M_1^2}.
\end{equation}
Numerically, this suppression due to the weak phase is at least about
$1/9$ due to the kinematic constraint,
\eqref{eq:kinematic-constraint}.  The requirement of a non-vanishing
strong phase implies that the asymmetry is large only in portion of
the Dalitz plot with distance of order $m_{H^\pm} \Gamma_{H^\pm}$ from
the point where the two resonances overlap. Thus, for the integrated
asymmetry there is an extra suppression of order
$\Gamma_{H^\pm}/m_{H^\pm}$.  When considering the fully integrated
asymmetry, we get an additional suppression due to the fact that the
phase space is nearly symmetric: $\Delta m_{\chi^\pm} \ll M_1$. This
suppression is not there for the sign weighted asymmetry. This hierarchy
gives a suppression of $\Delta m_{\chi^\pm}^2/M_1^2$. Putting these pieces together, we
can say that for order one CP-odd phase, the asymmetry in integrated
rates is roughly given by
\begin{equation}
  \label{eq:supp-tot-int}
  {\mathcal{A}}_{\rm CP}^{\rm int} \sim \frac{\Gamma_{H^\pm} \Delta
    m_{\chi^\pm}^2 |\mu M_2|}{m_{H^\pm} M_1^4},
\end{equation}
and the asymmetry in the weighted rates is roughly given by
\begin{equation}
  \label{eq:supp-tot-weigh} 
  \mathcal{A}_{\rm CP}^{\rm wgt} \sim
  \frac{\Gamma_{H^\pm} |\mu M_2|}{m_{H^\pm} M_1^2}.
\end{equation}
From these results we conclude that, in order to enhance the asymmetry
as much as possible, we would like to have the smaller parameters
$\mu$ and $M_2$ as close as possible to the larger parameter $M_1$
without cutting into phase space.

We now present some more specific results.  The tree-level
differential decay rate induced by the diagrams
Fig. \ref{fig:decay-charged-H} is given in Appendix
\ref{sec:diff-decay-rate}.  In order to study this decay rate, we
choose a specific point in MSSM parameter space.  We arbitrarily
parametrize the model such that 
the CP-violating phase is contained entirely in $\mu$ and the other
parameters are real.  The Bino mass $M_1$ is chosen to be much larger
than the other weak-scale masses so that there is sufficient phase
space to allow the relevant decay.  The other new dimensionful
parameters are chosen to be of order $100~{\rm GeV}$, but can be
varied in absolute scale without changing the results significantly.

\begin{table}[!ht]
  \centering
  \begin{tabular}{c @{\qquad} c}
    \hline\hline
    Parameter & Value \\
    \hline
    $M_1$ & $500~{\rm GeV}$ \\
    $M_2$ & $80~{\rm GeV}$ \\
    $\tan\beta$ & $5$ \\
    $M_{H_u}^2$ & $-(120~{\rm GeV})^2$ \\
    $M_{H_d}^2$ & $(250~{\rm GeV})^2$ \\
    ${\rm arg}(\mu)$ & $\pi/2$ \\
    \hline\hline
  \end{tabular}
  \caption{The choice of MSSM and soft SUSY-breaking parameters used
    to study CP-violation in the decays $\chi_4^0 \to \chi^\pm_1
    \chi^\mp_2 \chi^0_1$.   All other relevant parameters have
    been measured and are set to their values according to
    Ref.~\cite{Nakamura:2010zzi}.
  }
  \label{tab:param-choice}
\end{table}

\begin{figure}[!ht]
  \centering
\mbox{\subfigure{\includegraphics[width=.45\textwidth]{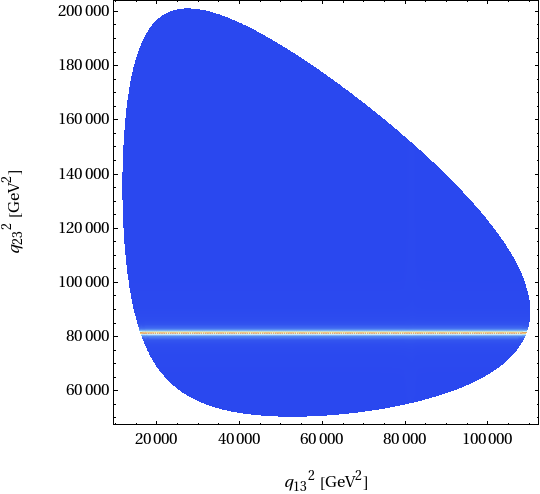}}\quad
  \subfigure{\includegraphics[width=.45\textwidth]{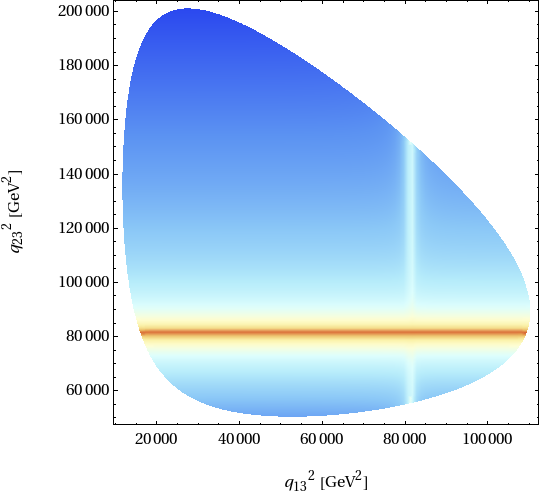} }}

\mbox{\subfigure{\includegraphics[width=.45\textwidth]{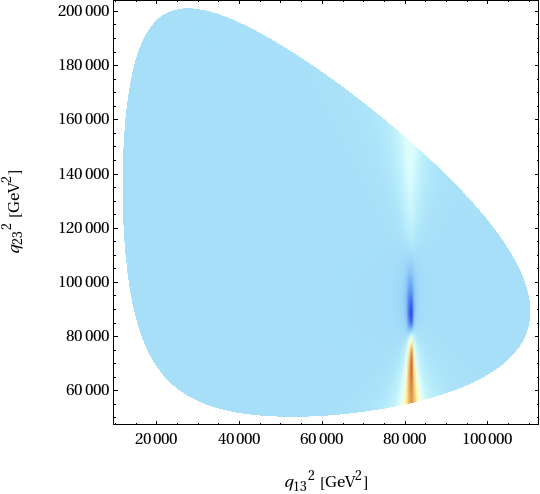}}\quad
  \subfigure{\includegraphics[width=.45\textwidth]{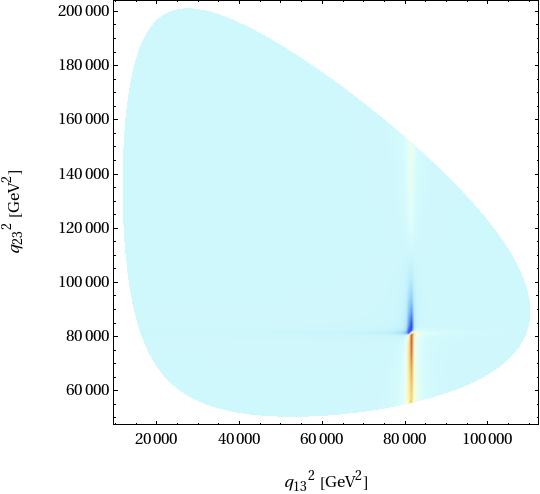} }}
  \caption{Dalitz plots for (a) the differential rate of $\chi_1^+ \chi_2^-
    \chi_1^0$ decay, (b) the log of the differential rate,  (c) the asymmetry $\mathcal{A}_\text{CP}$,  and (d) the significance
    $\frac{d\sigma_\text{CP}}{\sqrt{dq_{13}^2 dq_{23}^2}} \sqrt{\frac{\Gamma_{\chi^0_4}}{N}}$. The indices $1,2,3$ refer to $\chi_1^+, \chi_2^-$ and $\chi^0_1$ respectively.}
  \label{fig:dalitz-plots-mssm}
\end{figure}
The particular choice of parameters used for this study is given in
Table \ref{tab:param-choice}.  All other superpartners are assumed to be 
heavy or otherwise negligible.  Dalitz plots of the differential decay
rate, relative asymmetry, and significance as defined in
Sec.~\ref{sec:cp-even-phases} of the asymmetry for the
processes $\chi_4^0 \to \chi^\pm_1 \chi^\mp_2 \chi^0_1$, including
only the amplitudes involving a virtual charged Higgs, are 
shown in Fig.~\ref{fig:dalitz-plots-mssm}.
Many of the features that were obvious in the toy example are obscured
here due to the fact that the resonance in the $q_{13}^2$ direction
corresponding to the left diagram in Fig.\ \ref{fig:decay-charged-H} is
suppressed.  In fact using a linear color function for the Dalitz plot
(Fig.\ \ref{fig:dalitz-plots-mssm} (a)) this resonance is not even
visible and only the dominant one in the $q_{23}^2$ direction shows
up. In Fig.\ \ref{fig:dalitz-plots-mssm} (b) showing the log of the
differential decay rate also the $q_{13}^2$ resonance shows up, but is
suppressed by more than two orders of magnitude with respect to the
dominant one. The pattern of constructive and destructive interference
between the two resonances, which was clearly visible in the toy model
decay, is not visible from these figures, suggesting that the relevant
CP asymmetry is small. In Fig.\ \ref{fig:dalitz-plots-mssm} (c) we see
the resulting differential CP asymmetry which appears ``tilted''
towards the weak $q_{13}^2$ resonance with respect to the toy model
case. On top of this there is now a phase space 
dependence in the numerator of the amplitude due to the fact that the
external states are not scalars. We also observe the suppression due to the narrowness of the $H^\pm$ resonance, $\Gamma_{H^\pm}/m_{H^\pm} \simeq 0.5\%$.

Next, we calculate CP-violating integrated asymmetries.  As discussed
in Section \ref{sec:cp-even-phases}, an unweighted phase space
integration can be improved upon by introducing a relative sign
between the rates above and below 
the line $q_{13}^2 = q_{23}^2$ in phase space.  In particular, for the
scenario described in Table \ref{tab:param-choice}, the total rate
asymmetry is calculated to be $-3.5 \times 10^{-5}$, while introducing a relative sign
improves the asymmetry to $-6.5 \times 10^{-4}$.  The improvement is a
factor of almost $20$, roughly obtained by eliminating the suppression
$\Delta m_{\chi^\pm}^2/m_{\chi^0_4}^2 \sim 1/20$.

We have performed this study with the goal of demonstrating the
potential relevance of such CP violation to models of physics beyond
the Standard Model in general.  As such, we have worked from a bottom
up approach.  In particular, we have worked with a tree-level
SUSY Lagrangian with added soft SUSY breaking terms.  Renormalization
of the parameters from a UV SUSY breaking scheme can significantly
alter the spectrum and couplings.  Furthermore, a full study of this
scenario should include a UV theory of SUSY breaking that gives a
heavy, Bino-like neutralino at the weak scale.  On the other hand, so
long as the phase (\ref{eq:cp_vio_invariant}) is non-zero and the
decay studied is allowed to proceed on-shell, the strong phase due to
the virtuality of the intermediate charged Higgs will lead to a new
source of CP violation independent of the model's details.

A detailed study of collider prospects of this model is beyond the scope of
this paper.  We would, however, like to make a few remarks on what
would be necessary to observe the asymmetry in question. At the level
of theory, a more detailed study should also include the important
heavy Higgs contribution.  Experimentally, there are several
challenges that need to be overcome.  Quite generally, the significance of signal
depends on the following factors: the integrated luminosity, the
production cross section for $\chi_4^0$ and the branching ratio for
$\chi_4^0 \to \chi^\pm_1 \chi^\mp_2 \chi^0_1$, all affecting the 
number of events. Due to the smallness of the asymmetry in question
clearly a very large number of events is needed. Given a large
production of these decays, it is then necessary to identify the
events as having the correct structure. In the case where R-parity is
conserved, this issue is exacerbated by issues of combinatorics as the
heavy neutralino must be produced in conjunction with another
superpartner that could have similar decay modes. Our inability to
determine final state MSSM particle momenta ensures that we can only
determine the integrated asymmetry, which suffers from the additional
$\Delta m_{\chi^\pm}^2/m_{\chi_4^0}^2$ suppression. It might be
possible to circumvent this suppression by studying asymmetries in
kinematic observables, such as invariant masses of the SM decay
products of the decaying charginos.  We leave such investigations for
future work.

\section{Conclusions}\label{sec:conclusions}

The hope is that new physics will be soon discovered at the LHC. Once
it is discovered, we can turn to study all the parameters describing it. In
particular, we will study the masses, spins and couplings of the new particles. In
doing so, we would like to look for signals of CP violation.  
In this paper, we pointed out a new way to look for such signals:
looking for asymmetries in the Dalitz plots of cascade decays with 
unstable intermediate particles. The new observation is the fact that
a strong phases can arise even when there is only one intermediate
particle. This phase is present when there are two amplitudes in which the
intermediate particle has different virtuality.

This situation arises generally in cases where two
interfering diagrams exist with different orderings of the final
state particles. This effect can be present in both neutral and
charged particle decays. While we have elaborated mainly on neutral
particle decays in this paper, a study of a concrete example of CP
asymmetries in charged particle decays is in progress \cite{BBG2}.
The new observable we discuss is complementary to other observables that
have been discussed before, such as  triple product observables
and CP violation associated with oscillations.

It is particularly important to account for the kind of strong phase
described here in new physics models.  Most of these models, and
certainly ones in which we can hope to calculate any observables
reliably, are weakly coupled at sufficiently low energies.  In such
cases, a sizable strong phase can only come in processes involving
unstable particles. 
This could be important in essentially any beyond the SM scenario,
such as heavy neutrino decays, Kaluza-Klein state decays, and $W'$ or
$Z^\prime$ decays, to name just a few examples.  Even within the weak
sector of the MSSM, interference terms of the kind studied here are
relevant to the asymmetry in $\chi^0_i \to W^\pm H^\mp \chi^0_j$, 
where it is a subdominant contribution compared to chargino
mixing.  In many of these new physics scenarios, observables could be
complicated, as in the MSSM weak sector, by the existence of several
amplitudes for the decay, all with different strong phases.

The CP violating observable we have introduced could, in principle, be
relevant for SM physics as well, but it is not easy to come up with a
practical observable. In terms of fundamental particle decays, a
possible channel would in principle be $t \to  q_i q_j \bar q_k $ with
an intermediate $W$ exchange. However, the region where the $W$ is
approximately on shell in both diagrams lies outside the 
physical region of phase space. In addition, the weak phase is
highly suppressed due to the hierarchical structure of the CKM
matrix strongly favoring one decay channel over the other. We thus
have to rely on decays of composite particles, namely
hadrons. Indeed, there is a plethora of three body decays of $K$,
$D$ and $B$ mesons at our disposal. Kaon decays, however, do not
occur via a resonance. Neutral $D$ and $B_{d,s}$ meson decays are
not good examples either, since the same physics can be probed via
oscillations in a more effective way. We are thus left with charged
$B$ and $D$ meson decays. While it should be possible to find a
decay channel for which the intermediate particle can go on shell in
both diagrams, we are again confronted with the CKM hierarchy,
leading to a strong suppression of one channel with respect to 
the other in many cases. 

To conclude, most extensions of the SM include new
particles whose decays can lead to the type of CP violation that we
discuss here. Thus, we expect this type of CP violation to be relevant in
finding CP violating signals at the LHC and future colliders in many
of the possible scenarios for physics beyond the SM.

\subsection*{Acknowledgements}
This
work is supported by the U.S. National Science Foundation
through grant PHY-0757868 and CAREER award
PHY-0844667. MB would like to thank Andrzej Buras and the Technische Universit\"at M\"unchen for the hospitality during the completion of this work.


\begin{appendix}

\section{The Electroweak Sector of the MSSM}
\label{sec:electr-sect-mssm}

The charginos and neutralinos are the mass-basis superpartners of the
Higgs and Electroweak gauge bosons.  Their physics is determined by
three components of the Lagrangian:
\begin{itemize}
\item The superpotential
  \begin{equation}
    W = \mu H_u H_d
  \end{equation}
  leading to fermion terms
  \begin{equation}
    \mathcal{L} = - \mu \tilde{H}_u \tilde{H}_d + \hc
  \end{equation}
	
\item The supersymmetric gauge interactions
  \begin{multline}
    \mathcal{L} = - \sqrt{2} g \left( H_d^\dagger \frac{\sigma^a}{2}
      \tilde{H}_d \right) \tilde{W}^a - \sqrt{2} g \tilde{W}^a
    \left(\tilde{H}_d^\dagger
      \frac{\sigma^a}{2} H_d \right)  \\
    - \sqrt{2} g \left(H_u^\dagger \frac{\sigma^a}{2}
      \tilde{H}_u\right) \tilde{W}^a - \sqrt{2} g \tilde{W}^a \left(
      \tilde{H}_u^\dagger \frac{\sigma^a}{2} H_u \right)
    + \sqrt{2} g^\prime \left(H_d^\dagger \frac{1}{2} \tilde{H}_d\right) \tilde{B} \\
    + \sqrt{2} g^\prime \tilde{B} \left(\tilde{H}_d^\dagger
      \frac{1}{2} H_d \right) - \sqrt{2} g^\prime \left(H_u^\dagger
      \frac{1}{2} \tilde{H}_u\right) \tilde{B} - \sqrt{2} g^\prime
    \tilde{B} \left( \tilde{H}_u^\dagger \frac{1}{2} H_u \right)
  \end{multline}
	
\item The soft SUSY-breaking interactions
  \begin{equation}
    \mathcal{L} = - (M_1 \tilde{B}\tilde{B} + M_2 \tilde{W}^a
    \tilde{W}^a + b H_u H_d + \hc)
  \end{equation}
\end{itemize}

Notice that the Higgsino mass is determined only by the
superpotential, the gaugino mass only by the SUSY-breaking
interactions, and the mixing only by EWSB.  This structure means that
the mass difference between charginos will always be at least of order
$m_W$.  This fact will lead to a suppression of CP violating effects
in the chargino-neutralino sector, which are generally suppressed when
the mass difference is either much smaller or larger than some other
scale set by the width in the process.  

The resulting mass matrices after EWSB are \cite{Martin:1997ns}
\begin{equation}
  \label{eq:chargino-mass-mat}
  M_{\tilde{C}} = \begin{pmatrix}
    M_2 & \sqrt{2} s_\beta m_W \\
    \sqrt{2} c_\beta m_W & \mu \\
  \end{pmatrix}
\end{equation}
in the $(\tilde{W}^+,\tilde{H}_u^+),(\tilde{W}^-,\tilde{H}_d^-)^T$
basis and
\begin{equation}
  M_{\tilde{N}} = \begin{pmatrix}
    M_1 & 0 & -c_\beta s_W m_Z & s_\beta s_W m_Z \\
    0 & M_2 & c_\beta c_W m_Z & - s_\beta c_W m_Z \\
    -c_\beta s_W m_Z & c_\beta c_W m_Z & 0 & -\mu \\
    s_\beta s_W m_Z & - s_\beta c_W m_Z & -\mu & 0 \\
  \end{pmatrix}
\end{equation}
in the $(\tilde{B},\tilde{W}^0,\tilde{H}_d^0,\tilde{H}_u^0)$ basis.
These mass matrices are diagonalized by
\begin{equation}
  \label{eq:1}
  M_{\tilde{C}} = U^T M_{\tilde{C}}^{(D)} V,\qquad M_{\tilde{N}} = N^T
  M_N^{(D)} N,
\end{equation}
where $U$, $V$, and $N$ are unitary matrices.

The electroweak sector of the MSSM, including the Higgs fields, generically
violates CP symmetry with new physical phases.  We will now count
parameters and look at CP-violating invariants in this sector.  This
analysis has been done in ref. \cite{Dimopoulos:1995kn}, but we
reproduce it here with a different emphasis.

The electroweak sector has four complex parameters $M_i$, $\mu$, $b$
plus the real gauge couplings.  We would like to determine how many of
these parameters are physical.  Note that the electroweak sector
without potentials has a $U(1)_R \times U(1)_{PQ}$
global symmetry.  The symmetry is explicitly broken by the
superpotential and soft terms.  There is no residual symmetry in the
electroweak sector.  We are thus able to remove two of the four
complex phases in the parameters listed.  There are two remaining
physical phases in this sector.

\begin{table}[!ht]
  \centering
  \begin{tabular}{c @{\qquad} c @{\qquad} c}
    \hline\hline
    Field & $U(1)_R$ & $U(1)_{PQ}$ \\
    \hline
    $H_u$ & $1$ & $1$ \\
    $H_d$ & $1$ & $1$ \\
    \hline\hline
  \end{tabular}
  \caption{Charges of the (chiral) superfields under $U(1)_R \times U(1)_{PQ}$.}
  \label{tab:field_charges}
\end{table}
\begin{table}[!ht]
  \centering
  \begin{tabular}{c @{\qquad} c @{\qquad} c}
    \hline\hline
    Spurion & $U(1)_R$ & $U(1)_{PQ}$ \\
    \hline
    $M_i$ & $-2$ & $0$ \\
    $\mu$ & $0$ & $-2$ \\
    $b$ & $-2$ & $-2$ \\
    \hline\hline
  \end{tabular}
  \caption{Charges of the spurions under $U(1)_R \times U(1)_{PQ}$.}
  \label{tab:spur_charges}
\end{table}
Next, we would like to determine the invariants corresponding to these
phases.  For this, we perform a spurion analysis.  The charges of the
superfields under the symmetries are summarized in Table
\ref{tab:field_charges}.  After writing these down, the charges of the
spurions can be read off the potentials.  The $\mu$ term conserves $R$
charge, but violates $PQ$ symmetry.  In order to render that term
invariant, $\mu$ would need to have a charge of $-2$.  The gauginos
are invariant under $U(1)_{PQ}$, but they break $U(1)_R$ since they
are superpartners of the gauge bosons, which must have $R$ charge 0.
The gauginos have $R$ charge $1$, so the gaugino masses have a spurious $R$
charge of $-2$.  Finally, the $b$ term violates both symmetries.
$U(1)_{PQ}$ is violated as in the $\mu$ term, so $b$ has the same $R$
charge as $\mu$.  It also violates $U(1)_R$ since it should have $R$
charge 0, not 2 as in the superpotential.  $b$ must then have a
spurious $R$ charge of 2.  The charges of
the spurions are summarized in Table \ref{tab:spur_charges}.

All observables must be proportional to Hermitian combinations
of parameters that have 0 spurious charge.  CP violating observables
should be proportional to the imaginary part of combinations with 0
spurious charge.  The imaginary part vanishes in the CP conserving
case and renders the combination real.  There are two classes of such
observables in the current case.  The first is the class of
observables formed out of gaugino masses alone: ${\rm Im}(M_1^*
M_2)$.  The other class of observables involves $\mu$.  Such
observables must also involve $b$ since it is the only other spurion 
with $PQ$ charge.  In particular, we must use the combination $\mu^*
b$, which has no $PQ$ charge but has $R$ charge $-2$.  To form an
invariant we must include one of the gaugino masses.  However, the
two possible such terms (one for each gaugino mass) are not
independent since they can be written in terms of just one of the
possible combinations, as well as combinations of only gaugino masses.
In what follows, we will discuss only the electroweak sector.  We will
further neglect mixing with Bino for simplicity.  This approximation
is justified when the mass $M_1$ is much larger than $M_2$, $\mu$ and
$b$.  Then, the only relevant CP violating invariant to study is
\begin{equation}
  \label{eq:cp_vio_invariant}
  {\rm Im}(\mu^* b M_2^*).
\end{equation}
While there are generally strong bounds on this phase due to the
non-observation of electric dipole moments \cite{Nakamura:2010zzi}, the
bounds are model dependent.  They come from loops 
involving the sleptons \cite{Ibrahim:1998je,Brhlik:1998zn}.  We assume
that we can make these loops small by, for example, making the
sleptons very heavy, so that the region of parameter space we will
study is not excluded by indirect measurements.

\newpage

\section{Differential Decay Rate of Heavy Neutralino}
\label{sec:diff-decay-rate}

We use the notation of Ref.\ \cite{Dreiner:2008tw}.  The differential
decay rate of the heavy neutralino via the diagrams 
Fig. \ref{fig:decay-charged-H}, $\chi_4^0\to \chi_1^+ \chi_2^-\chi_1^0$, is given by
\begin{eqnarray}
  \label{eq:diff-decay-rate-mssm}
  \frac{d\Gamma}{dq_{13}^2 dq_{23}^2} & = & \frac{1}{(2 \pi)^3}
  \frac{1}{32 m_0^3} \nonumber\\
  ~ & ~ & \left[\frac{1}{(q_{13}^2 -
    m)^2 + \Gamma^2 m^2} \left[(m_0^2 +
    m_2^2 - q_{13}^2) (|\lambda_{02}^+|^2 + |\lambda_{02}^-|^2) + 4
    m_0 m_2 {\rm Re}(\lambda_{02}^+ \lambda_{02}^-)\right]\right. \nonumber\\
  ~ & ~ & \left[(q_{13}^2 - m_1^2 - m_3^2) (|\lambda_{31}^+|^2 + |\lambda_{31}^-|^2) - 4
    m_1 m_3 {\rm Re}(\lambda_{31}^+ \lambda_{31}^-)\right] + (1
  \leftrightarrow 2) + \nonumber
  \\
   ~ & ~ & 2 {\rm Re}\bigg\{\frac{1}{(q_{13}^2 - m^2 + i m \Gamma)
       (q_{23}^2 - m^2 - i m \Gamma)} \times  \nonumber\\
   ~ & ~ & \quad \left[m_0 m_1 (q_{23}^2 - m_2^2
       -m_3^2) (\lambda_{32}^- \lambda_{01}^{-*} \lambda_{02}^{-*}
       \lambda_{31}^{+*} + [+ \leftrightarrow -]^*) + \right.\nonumber\\
       ~ & ~ & \qquad (q_{13}^2 q_{23}^2 - m_0^2 m_3^2 - m_1^2 m_2^2) (\lambda_{01}^+
     \lambda_{32}^- \lambda_{02}^{-*} \lambda_{31}^{+*} + [+
     \leftrightarrow -]^*) + \nonumber\\
    ~ & ~ &  \qquad  m_1 m_2 (q_{13}^2 + q_{23}^2 - m_1^2 - m_2^2) (\lambda_{02}^+
   \lambda_{32}^- \lambda_{01}^{-*} \lambda_{31}^{+*} + [+
   \leftrightarrow -]^*) + \nonumber \\
   ~ & ~ &  \qquad m_0 m_3 (q_{13}^2 + q_{23}^2 - m_0^2 - m_3^2) (\lambda_{31}^-
 \lambda_{32}^- \lambda_{01}^{-*} \lambda_{02}^{-*} + [+
 \leftrightarrow -]^*) + \nonumber\\
~ & ~ & \qquad m_0 m_2 (q_{13}^2 -m_1^2 -m_3^2) (\lambda_{01}^+
\lambda_{02}^+ \lambda_{32}^- \lambda_{31}^{+*} + [+ \leftrightarrow
-]^*) + \nonumber\\
~ & ~ & \qquad m_1 m_3 (q_{13}^2 - m_0^2 -m_2^2) (\lambda_{01}^+
\lambda_{31}^- \lambda_{32}^- \lambda_{02}^{-*} + [+ \leftrightarrow
-]^*) + \nonumber\\
~ & ~ & \qquad  m_2 m_3 (q_{23}^2 - m_0^2 -m_1^2) (\lambda_{02}^+
\lambda_{31}^- \lambda_{32}^- \lambda_{01}^{-*} + [+ \leftrightarrow
-]^*) -\nonumber\\
~ & ~ & \qquad \left. 2 m_0 m_1 m_2 m_3 (\lambda_{01}^+ \lambda_{02}^+
\lambda_{31}^- \lambda_{32}^- + [+ \leftrightarrow -]^*)\right]\bigg\}\Bigg], 
\end{eqnarray}
where $m = m_{H^+}$, $\Gamma = \Gamma_{H^+}$, $m_0 = m_{\chi^0_4}$,
$m_1 = m_{\chi^+_1}$, $m_2 = m_{\chi^-_2}$, $m_3 = m_{\chi^0_1}$, and
$\lambda_{ij}^\pm = Y^{H^\pm \chi^0_{a(i)} \chi^\mp_{j}}$ with
$a(0) = 4$, $a(3) = 1$, and $j=1,2$.  The notation $[+
\leftrightarrow -]^*$ means exchange $\lambda^\pm \leftrightarrow
\lambda^{\mp *}$ with the same indices. The differential decay rate for $\chi_4^0 \to \chi_1^- \chi_2^+ \chi_1^0$ can be obtained from \eqref{eq:diff-decay-rate-mssm} by interchanging the indices 1 and 2 at all places.

\end{appendix}


\end{document}